\documentclass{emulateapj}

\newcommand{\etal}{et~al.\ }

\slugcomment{To Appear in The Astrophysical Journal}

\shorttitle{Far-IR Photometry of the TWA}
\shortauthors{Low \etal}

\begin{document}

\title{Exploring Terrestrial Planet Formation in the TW Hydrae Association}
\vskip 0.2in

\author{Frank J. Low and Paul S. Smith}
\affil{Steward Observatory, The University of Arizona,
    Tucson, AZ 85721; flow@as.arizona.edu, psmith@as.arizona.edu}
\vskip 0.1in

\author{Michael Werner and Christine Chen{\altaffilmark{1,}\altaffilmark{2}}}
\affil{Jet Propulsion Laboratory, 4800 Oak Grove Dr., Pasadena, CA  91109; mww@ipac.caltech.edu}
\altaffiltext{1}{National Research Council Resident Research Associate.}
\altaffiltext{2}{Current address: National Optical Astronomy Observatories,
Tucson, AZ  85719; cchen@noao.edu}
\vskip 0.1in

\author{Vanessa Krause}
\affil{Division of Physics, Mathematics, and Astronomy, California Institute of Technology, MS 103-33, Pasadena, CA  91125; krause@caltech.edu}
\vskip 0.1in

\author{Michael Jura}
\affil{Department of Physics and Astronomy, University of California, Los Angeles, Los Angeles, CA  90095; jura@clotho.astro.ucla.edu}
\vskip 0.1in

\author{Dean C. Hines}
\affil{Space Science Institute, 3100 Marine Street, Suite A353,Boulder, CO  80303}
\vskip 0.2in

\begin{abstract}

{\it Spitzer Space Telescope\/} infrared measurements are presented
for 24 members of the TW~Hydrae association (TWA).
High signal-to-noise 24~$\mu$m photometry is presented for all of these stars,
including 20 stars
that were not detected by {\sl IRAS\/}.
Among these 20 stars, only a single object, TWA~7, shows excess
emission at 24~$\mu$m and at the level of only 40\% above the
star's photosphere.
TWA~7 also exhibits a strong 70~$\mu$m excess that is a factor of 40
brighter than the stellar photosphere at this wavelength.
At 70~$\mu$m, an excess of similar magnitude is detected for TWA~13,
though no 24~$\mu$m excess was detected for this binary.
For the 18 stars that failed to show measurable IR excesses, the sensitivity
of the current 70~$\mu$m observations does not rule out substantial cool
excesses at levels 10--40$\times$ above their stellar continua.
Measurements of two T~Tauri stars, TW~Hya and Hen~6-300, confirm that their
spectacular IR spectral energy distributions (SEDs) do not turn over even by
160~$\mu$m, consistent with the expectation for their active accretion disks.
In contrast, the {\it Spitzer\/} data for the luminous planetary debris
systems in the TWA, HD~98800B and HR~4796A,
are consistent with single-temperature blackbody SEDs and agree
with previous IR, sub-millimeter, and millimeter measurements.
The major new result of this study is the dramatic bimodal distribution found
for the association in the form of excess emission at a
wavelength of 24~$\mu$m, indicating
negligible amounts of warm ($\gtrsim 100$~K) dust and debris around 20
of 24 stars in this group of very young stars.
This bimodal distribution is especially striking given that the four 
stars in the association with strong IR excesses are $\gtrsim 100\times$
brighter at 24~$\mu$m than their photospheres.
Clearly, two terrestrial planetary systems, HD~98800B and HR~4796A,
exist in some form.
In addition, there are at least two
active accreting objects, TW~Hya and Hen~6-300, that may still be
forming planetesimals.
The remaining stars may possess significant amounts of cold dust, as
in TWA~7 and TWA~13, that have yet to be found.

\end{abstract}


\section{Introduction}

While calibrating the {\it Infrared Astronomical Satellite\/} ({\sl IRAS\/})
\citet{aumann84} and \citet{gillett86}
discovered large far-infrared
excesses around four nearby stars: Vega,
Fomalhaut, $\beta\/$~Pictoris, and $\epsilon\/$~Eridani.
Release of the {\sl IRAS\/} Point Source Catalog
led investigators to many other young, nearby dwarf stars displaying
excess emission at mid- and far-IR wavelengths
\citep[e.g.,][]{walker88,reza89,hetem92,mannings98,zuckerman93}. 
As a result, numerous investigations of these objects
from the ground have added critical insights concerning circumstellar material
\citep[e.g.,][]{jayawardhana98,koerner98,kenyon02,kenyon04}.
The {\it Hubble Space Telescope\/} ({\sl HST\/}) and
the {\it Infrared Space Observatory\/} ({\sl ISO\/})
\citep[][and references therein]{habing01,spangler01,meyer00}
have also given us key insights into specific objects, including images taken
in the near-IR showing scattered light from thin disks
of dust that are shown to absorb starlight and re-emit at much longer
wavelengths \citep[e.g.,][]{schneider99}. 
Thus, in the last 20 years a body of
observations, measurements, analysis and theory
have accumulated that leave no doubt about
the reality of terrestrial planetary material around
young stars.
Now, investigations of planetary debris systems (PDSs)
are underway with the new
tools provided by the {\it Spitzer Space Telescope\/} \citep{werner04}.
 
We have initiated a program
using {\it Spitzer\/} to
explore a unique and varied sample of late-type stars that are
nearby, $\sim$40--100~pc, and show one or more indications of young age.
Our total sample of over 120 stars has been prioritized into
three groups: (1) the first 19 stellar systems identified
by common space motion
as members of the
TW~Hydrae association \citep[TWA;][]{kastner97,webb99},
(2) $\sim$60 stars that show significant lithium absorption
indicating young ages, noting that all of the TWA systems easily meet this
requirement and therefore overlap this sample,
(3) a group of $\sim 60$ stars with less certain ages based on either
detection of X-rays or H$\alpha$ in emission.

A combination of special circumstances dictate that we start our
exploration with the TW~Hydrae association.
First, the estimated age of the association ranges
from 8--10 million years (Myr) and the ages of
these stars are essentially equal \citep{stauffer95}.
Second, the TWA members
are among the closest young stars available.
Third, most association members lie in excellent, low-background regions of the
sky for far-IR observations. 
{\sl IRAS\/} has already shown that four of
the brightest IR excesses known are found in the TWA.
Additional
important factors are that five of the TWA stars have {\sl Hipparcos\/}
distances
averaging around 50~pc, their spectral types range from A0V to M5V with most
objects being of M or late K spectral type,
and almost all members show signs of
T~Tauri activity. 
TW~Hydrae itself has been classified as a classic T~Tauri system despite
its apparent isolation \citep{herbig78,rucinski83}. 
In contrast, and
within the same association, the brightest and most dramatic PDS,
HD~98800, shows no evidence for current
accretion of primordial material
onto the star and has served as the most direct proof of ongoing
terrestrial planet formation \citep[e.g.,][]{low99}.

\section{Observations}

Observations of 24 members of the TW Hydrae association
were made using {\it Spitzer\/}.
These include 14 single stars and binaries that are unresolved at
24~$\mu\/$m and five binary systems at least partially resolved by
{\it Spitzer\/} at 24~$\mu\/$m.
The Multi-band Imaging Photometer for {\it Spitzer\/}
\citep[MIPS;][]{rieke04} 
provided measurements or upper limits of the flux densities ($F_\nu\/$) at
24~$\mu\/$m and 70~$\mu\/$m
for the entire sample.
MIPS 160~$\mu\/$m photometry was also obtained for the four IR-excess
systems discovered by {\sl IRAS\/} (i.e., TW~Hya, Hen~3-600, HD~98800B, and
HR~4796A).
The observations were made between 2004 January 30 and June 21, with each
target observed in all of the intended wavelength bands during a single
visit by {\it Spitzer\/}.
Table~1 lists the stars in the association observed with MIPS along with basic
data from previous measurements.
Six other systems have been identified as members of the TWA
\citep[see e.g.,][]{song03,webb00a} since the final planning of the
{\it{Spitzer\/}}/MIPS observations, but were not included in the target list 
because of time limitations.

The MIPS data are reduced using the data analysis tool (DAT) of the MIPS
instrument team \citep{gordon05}.
The images are corrected for distortion, registered, and mosaicked
to compensate for the movements of the spacecraft and cryogenic
scan mirror mechanism (CSMM) within MIPS that are employed
to produce a series of spatially dithered images on the three detector arrays.
Aperture photometry is then performed on the calibrated, mosaicked images.
Table~2 summarizes the MIPS bandpasses, aperture photometry parameters,
and the calibration factors adopted for the data reduction.
The data have been color corrected using the relevant tables in 
the {\it Spitzer\/} Observer's Manual.

The 24~$\mu\/$m data were collected using the standard MIPS point source
observing template that obtains 14 images of the target at
various standard locations on the 128$\times$128-pixel Si:As
detector array during
each cycle of the template.
Either 3~s or 10~s integration times and 1--4 complete template cycles
were used to ensure high signal-to-noise ratio (S/N) measurements of
the targets.
Total observation times were estimated for each target so that
a S/N$ > $10 measurement was achieved
at the expected 24~$\mu\/$m flux level of the star's photosphere,
and all stars were detected to at least this S/N. 
Indeed, the median S/N is $\sim$160, though the internal photometric
calibration is estimated to be 1--2\% in this bandpass
\citep[e.g.,][]{rieke05}.

For the aperture photometry of the 24~$\mu\/$m data,
an aperture 15\arcsec\ in radius ($r\/$) was typically used
along with a sky annulus centered on the aperture and having
an inner radius of $\sim$63\arcsec\ and width of 25\arcsec.
The large sky annulus was chosen to avoid as much flux as reasonable from
the {\it Spitzer\/} plus MIPS point spread function (PSF) of the target.
A larger aperture ($\sim$37\farcs 5 radius) was used for four
binary TWA systems that are at least
partially resolved at 24~$\mu\/$m (TWA~8, 9, 13, and 15).
The relative brightnesses of individual components of the binaries were
estimated by comparing the fluxes measured in $r = 2$\farcs 5
apertures centered 
on the components.
For TWA~9 and 13, a third source of 24~$\mu\/$m flux fell within the large 
photometric apertures.
The flux from these sources were subtracted and are not included 
in the values for $F_\nu\/$ at 24~$\mu\/$m.

Only HD~98800 was bright enough that the core of the PSF saturated 
even in the minimum observation time allowed at 24~$\mu\/$m.
In this case, the flux in the saturated core was estimated by measuring
the flux in the wings of the PSF.
This was done for the entire sample as well. 
The exercise on the non-saturated stars showed that the 24~$\mu\/$m flux
in the saturated portion of the PSF for HD~98800 could be recovered
to within 5\%.
After correcting for the saturated core, the 24~$\mu\/$m flux density
was measured for HD~98800 in the usual manner.

Observations, data reduction, and aperture photometry for the 70~$\mu\/$m
data are similar to those at 24~$\mu\/$m.
However, the pixel scale of the 70~$\mu\/$m
wide-field optical train
($\sim$10\arcsec~pixel$^{-1}$) dictates that none of the binaries in the TWA
can be resolved except for TWA~19.
Therefore, we have used a uniform photometry aperture (3-pixel radius) and
sky annulus (radius = 4--6~pixels) for the stars observed in the 70~$\mu\/$m
wide-field mode.
The chosen photometry aperture is very large compared to the
$\lesssim 1$\arcsec\ pointing accuracy of the telescope, eliminating
the ambiguity in both the target position on the array and the determination
of flux density upper limits for undetected sources. 

TW~Hya, HD~98800, and HR~4796A
were observed in the 70~$\mu\/$m narrow-field mode
to prevent saturation by these bright IR sources as this optical train
halves the pixel scale.
For these observations, an $r = 5$\farcs 67 aperture was chosen.
Instead of an annulus centered on the target to estimate the sky
contribution within the photometric aperture, rectangular regions on either 
side of the target were utilized because of the narrow, 16-pixel width
of the usable
portion of the detector array.

Because of the degraded sensitivity and higher instrumental noise of the  
70~$\mu\/$m Ge:Ga detector array relative to pre-launch estimates,
it was not practical to obtain measurements accurate enough to detect
objects at the expected levels of the TWA photospheres.
As a result, the 70~$\mu\/$m observations were tailored to detect excesses
at flux levels $\gtrsim$10$\times$ the expected 70~$\mu\/$m photospheric flux 
density with a reasonably high degree of confidence ($>$3--4$\sigma\/$).
Unfortunately, most of the TWA observations occurred before the bias
level for the 
70~$\mu\/$m detector was reduced.
During the period before the bias change, the overall noise in the detector
was greater and there were more pronounced changes seen in background levels
across array modules.
Because of these instrumental effects, a conservative criterion of
5$\sigma\/$ above background was adopted 
for a detection at 70~$\mu\/$m.

All four targets observed at 160~$\mu\/$m were detected at a S/N ranging
from $\sim$6--160 using the standard MIPS photometry observing template.
In all cases, an aperture of $r = 64$\arcsec\ was used for the photometry.
The sky contribution within the photometric aperture was estimated using 
rectangular regions on either side of the target.
As with the 24~$\mu\/$m and 70~$\mu\/$m data, the 160~$\mu\/$m photometry is 
corrected for the loss in flux caused by the use of apertures of finite 
size not encompassing the total flux from the PSF.
The multiplicative aperture corrections to a
hypothetical aperture of infinite extent
are listed in Table~2.

The 160~$\mu\/$m data were also corrected for the blue filter leak discovered 
in this bandpass after the launch of {\it Spitzer\/}.
This leak allows light at $\sim$2~$\mu\/$m in wavelength to reach the detector
which is sensitive to near-IR radiation.
Tests made during flight have shown that
the contaminating flux is about 15$\times$
the flux received from a star's photosphere at 160~$\mu\/$m assuming a 
Rayleigh-Jeans law between the near and far-IR.
In all four cases in the TWA where 160~$\mu\/$m measurements were made, the
expected photospheric 160~$\mu\/$m flux density is $\ll$1\% of the measured signal.
Given the bright far-IR excesses of these stars, the filter leak corrections
are small.
 
In addition to broad-band photometry, observations of TW~Hya and HD~98800
were made using MIPS in its SED mode.
These observations were made on JD~=~2453395, roughly a year after the 
MIPS photometry of TW~Hya was obtained and about seven months after the 
photometry of HD~98800.
In this observing mode, the CSMM brings light from the telescope to a reflective
diffraction grating that also performs the functions of a slit and 
collimator.
The collimated, dispersed light is then focused onto the 70~$\mu\/$m
detector array.
The result is a very low-resolution ($R = \lambda/\Delta\lambda \sim 15$-25)
first-order spectrum covering a wavelength range of 52--97~$\mu\/$m.
The slit/grating has a width of two pixels ($\sim 20$\arcsec) and the
dispersion is 1.7~$\mu\/$m~pixel$^{-1}$.
An unusable portion of the 70~$\mu\/$m array restricts the slit
length to $\sim 2$\farcm7.
Also, a dead 4$\times$8-pixel readout at one
end of the slit further reduces the effective length of the slit where
the full spectral range can be sampled.

Observations in SED mode involve chopping the CSMM between the
target and a region 1--3\arcmin\ away to sample the background, and
using small spacecraft moves to dither the target between two
positions along the slit.
Data reduction is again handled by the DAT in essentially the same manner
as for 70~$\mu\/$m broad-band imaging.
The result is co-added ``on-'' and ``off-target'' 
images.
After the background image has been subtracted from the target, the
spectrum is extracted using an extraction aperture of set width (8~pixels).
Flux calibration of the TW~Hya and HD~98800 spectra was accomplished
using a MIPS SED-mode observation of $\alpha$~Boo made on JD~=~2453196.
The 50--100~$\mu\/$m continuum of $\alpha$~Boo is assumed to be 
$\propto \lambda^{-2}$ and $F_\nu = 14.7$~Jy at the effective wavelength of
the 70~$\mu\/$m filter bandpass.
The observations and spectral extractions of $\alpha$~Boo and the
two TWA systems were identical
and the same DAT calibration files were used for every target.

\section{Results}

Table~3 summarizes the MIPS photometric results for the TWA.
Several stars in the association are known optical variables and we give the
epoch of the start of the observations in the second column of the table.
The 24~$\mu\/$m, 70~$\mu\/$m, and 160~$\mu\/$m flux densities are given in units
of milli-Janskys (mJy; where
1~mJy~= $10^{-26}$~erg~cm$^{-2}$~s$^{-1}$~Hz$^{-1}$).
Uncertainties in the flux densities are the square root of the
quadratic sum of the
measurement and calibration uncertainties,
noting that most relative uncertainties are much lower.
Also listed in Table~3 is the S/N of the detections in each MIPS band.
Although some of the S/N values are extremely high, the uncertainties quoted
for the photometry
in all three bandpasses are dominated by uncertainties in the absolute flux
calibration of the instrument.
These are estimated to be $\sim$10\% for the 24~$\mu\/$m band and $\sim$20\%
for the 70~$\mu\/$m and 160~$\mu\/$m bandpasses.
Upper limits assigned in Table~3 to non-detections at 70~$\mu\/$m are at
the level of 3$\sigma\/$ over the noise of the background.

Our primary goal for this observational program is to detect
and measure IR excesses in the TWA.
This requires that good estimates of $F_\nu\/$ be made for
the TWA stellar photospheres at the 
effective wavelengths of the MIPS bandpasses.
For all of stars in the sample, stellar photospheres were derived
using a grid of Kurucz \citep{kurucz79} and ``NextGen''
\citep{hauschildt99} photospheric models and fitting available optical 
photometry combined with near-IR
flux densities compiled from the Two-Micron All Sky Survey (2MASS)
Point Source Catalog.
Temperature is a free parameter in the fitting, but in all cases
solar metallicity and $\log {\rm g} = 4.5$ are assumed.
The NextGen models are known to yield better fits to
late-type main sequence stars than Kurucz models,
and were used for all TWA members except for HR~4796A and TWA~19A, which
have much higher effective temperatures than the rest of the sample.
The best-fitting model temperatures ($T_*\/$)
of the TWA are listed in Table~1.

\subsection{24~$\mu\/$m Excesses}

Comparison of the 24~$\mu\/$m data for the TWA with the
2MASS $K_s\/$-band flux densities
of these stars
shows that the association can essentially
be divided into two populations.
Figure~1 plots the 24-to-2.17~$\mu\/$m flux density ratio against the
flux density in the $K_s\/$-band.
In all cases, the 2MASS photometry is consistent with the 
stellar photospheres emitting the near-IR light.
The four stars with IR excesses discovered by {\sl IRAS\/}
are about a factor of 100 brighter at 24~$\mu\/$m relative to $K_s\/$
than the other stars in 
the sample.
In fact, for the stars with flux ratios $<$0.02, the data are consistent
with the 24~$\mu\/$m flux being solely from the photosphere. 
As a guide, a dashed line in Figure~1 marks the limit of
$F\/$(24~$\mu\/$m)/$F_{K_s}\/$ if both bandpasses fall within the 
Rayleigh-Jeans regime of the stellar continuum.
Since the TWA members are typically spectral type K and M,
the $K_s\/$ band falls
close to the emission peak of the photosphere.
As a result, the stars typically
lie above the Rayleigh-Jeans limit in Figure~1.

Figure~2 further explores the empirical relation between the 24~$\mu\/$m
flux density with the near-IR.
The stars with previously known IR excesses or
$T_* > 4500$~K
are omitted from Figure~2,
and the flux ratio shown in Figure~1 is now plotted against the temperature
of the model stellar photosphere.
The majority of the 19 stars plotted form a locus that is
offset from the $F\/$(24~$\mu\/$m)/$F_{K_s}\/$ ratios expected from
blackbodies.
Spectra derived from the NextGen models generally give a good
approximation to the observed flux ratios, although the TWA
members have not reached the main sequence.
The unweighted average flux ratio
$\langle F({\rm 24}\mu{\rm m})/F_{K_s}\rangle = 0.015$ with an
RMS of only 0.004.
The concentration of the points
close to the expected values 
from the NextGen models in Figure~2
suggests that we are
seeing no IR excess at 24~$\mu\/$m for most of these
late-type, young stars.
The MIPS photometry is consistent with the 24~$\mu\/$m flux
emanating from the photosphere for most of the members of the TWA.
One example is TWA~6, where the MIPS 24~$\mu\/$m measurement supports
the contention by \citet{uchida04} that the 5--20~$\mu\/$m spectrum
of this star taken with {\it Spitzer\/} and the Infrared
Spectrograph \citep[IRS;][]{houck04} is indeed the spectrum of the
photosphere.

Although the NextGen models yield a much better brightness
estimate of the 
photosphere at 24~$\mu\/$m than blackbodies in this temperature range,
Figure~2 shows that the flux ratio is possibly
more dependent on the model temperature than is actually 
observed.
Note, TWA 12, 13B, and 17 fall
well below the NextGen curve.
These deviations and the intrinsic scatter of the data are intriguing, but the
limited number of objects in the sample complicate a detailed investigation
of their possible causes, especially considering that
only temperature was varied in constructing the models and that
the TWA stars are younger
than the nominal zero-age main sequence stars described by the NextGen models.
In the case of TWA~17, the 24~$\mu\/$m flux density is measured to be
only $1.5 \pm 0.2$~mJy 
and the source is detected at a $S/N = 13$ with MIPS.
Therefore, we do not confirm the possible
excess at 12 and 18~$\mu\/$m reported from ground-based observations
by \citet{weinberger04}.

The measurement of $F\/$(24~$\mu\/$m)/$F_{K_s}\/$ for TWA~7
suggests that
it possesses a 24~$\mu\/$m excess that contributes $\sim$40\% to 
the total brightness of this system at this wavelength.
Except for the four {\sl IRAS\/} detections, TWA~7 has a
higher flux ratio in Figures~1 and 2 than all of the other TWA members.
The next highest ratio belongs to TWA~9B,
but its deviation from the expected flux of the photosphere is less than 
half of that shown by TWA~7, and is no larger than the flux deficits at
24~$\mu\/$m measured for TWA~12, 13B, and 17.
Since the 2MASS and MIPS measurements are not simultaneous,
an alternate explanation for the apparent IR excess in TWA~7 could be
the variability of
the star which is not uncommon for T~Tauri stars, although there
are no published data suggesting that this star is variable.
Also, the clustering of most measurements around
$\langle F\/$(24~$\mu\/$m)/$F_{K_s}\rangle\/$ is a clear demonstration that
most stars in the TWA do not, in fact, exhibit significant
variability.

\subsection{70~$\mu\/$m Excesses}

Regardless of the strength of a possible 24~$\mu\/$m excess in the
observed continuum of TWA~7, this object is detected at 70~$\mu\/$m
at a flux level $\sim$40$\times$ brighter than the photospheric
emission estimated for the star at this wavelength.
Figure~3 presents the ratio between the observed and
predicted 70~$\mu\/$m flux densities.
Only six of the 20 systems (TWA~19 is resolvable at 70~$\mu\/$m) are 
detected with MIPS at 70~$\mu\/$m.
TWA~13 joins TWA~7 and the four {\sl IRAS\/} sources in having a detected
70~$\mu\/$m excess.
TWA~13 cannot be resolved with {\it Spitzer\/} at
70~$\mu\/$m, so it is unknown which of the two components
produces the excess, or if both components of the binary contribute
to the observed flux.

Figure~3 shows the same basic division in the TWA sample as
is found at 24~$\mu\/$m.
The 70~$\mu\/$m excesses for TWA~7 and 13 are $\gtrsim$10$\times$ 
fainter than the other detections.
One-sigma upper limits are also shown in Figure~3 for the remainder
of the sample.
As discussed in \S2, the sensitivity limits of the 70~$\mu\/$m array 
render detection of the photospheres of the TWA sample impractical.
Excesses similar to those of TWA~7 and 13 cannot be ruled out 
for many members of the association, and it is clear from the
data that it is not necessary for an object to show an 
excess at 24~$\mu\/$m to find one at 70~$\mu\/$m. 
For instance, there is no hint from the 24~$\mu\/$m
measurement of TWA~13 of an excess at longer wavelengths.
There are three stars where a 70~$\mu\/$m excess at
$>$10$\times$ the flux density of the photosphere can 
be ruled out at the 3$\sigma\/$ confidence level: TWA~2, 5A, and 8A.
Therefore, the minimum range in strength for 70~$\mu\/$m excesses in the
TWA is greater than a factor of 300.

\subsection{Spectral Energy Distributions}

The SEDs of the six systems in the TWA
with confirmed IR excesses are shown in Figure~4.
It is readily apparent that this small sample of young stars exhibits
a rich variety of IR properties.
The T~Tauri stars TW~Hya and Hen~3-600 show no hint of their SEDs turning 
over even out to 160~$\mu\/$m.
In fact, the MIPS 160~$\mu\/$m measurement for TW~Hya reveals that the
spectrum of the presumed optically thick accretion disk
continues to rise into the infrared.
Similar to TW~Hya, the MIPS data for Hen~3-600 are consistent
with the earlier {\sl IRAS\/}
results and show that the 70 and 160~$\mu\/$m flux 
densities are roughly equivalent.

In contrast, HD~98800A and B do not show the extreme activity associated with
T~Tauri stars.
However, HD~98800B generates a huge IR excess that is well fit by a
single blackbody
with $T = 160$-170~K \citep{low99,hines04}, whereas component A lacks an easily measured
excess even though the total luminosities of the two companions are
virtually identical.
The MIPS photometry is consistent with earlier results.
Likewise, the IR excess of HR~4796A produces an IR SED that can be fit
by a single blackbody, but it is significantly cooler than found for HD~98800B
\citep[$T = 110$~K;][]{jura93}.
With the addition of the {\it Spitzer\/} observations, we fit this
excess with a $T = 108$~K blackbody.
Given that HR~4796A is an A0V star, it is not surprising that it
has an IR excess because of the finding that younger A stars have a
higher probability of possessing a 24~$\mu\/$m excess and having stronger 
thermal excess emission than older A stars \citep{rieke05}.
It is of note that HR~4796A has by far (by $\sim$10$\times$)
the largest ratio of 24~$\mu\/$m excess
emission-to-photospheric emission of the 265 A and B stars
examined by \citet{rieke05}.

Figure~5 presents the MIPS SED-mode observations of TW~Hya and HD~98800B.
These low-resolution spectra do not show any spectral features of high
equivalent width in either object.
Arcturus was used to calibrate the spectra and they are consistent with
the 70~$\mu\/$m flux densities derived from the photometry.
The IR spectral slopes of TW~Hya and HD~98800B are quite different, with
the continuum of TW~Hya being much redder than that of HD~98800B.
A power-law having a spectral index, $\alpha\/$, 
where $F_\nu \propto \nu^\alpha\/$, is adequate to describe these data.
For TW~Hya, $\alpha = -0.33 \pm 0.04$, and
is consistent with the finding that the
160~$\mu\/$m flux density is higher than the 70~$\mu\/$m
measurement.
The slope of a single power-law fit to the continuum of HD~98800B between 52 and
97~$\mu\/$m yields $\alpha = 1.12 \pm 0.04$.
The blackbody fit with $T_{BB} = 160$~K to the PDS of HD~98800B
is also shown in Figure~5, and its 
slope within this spectral region is in good agreement with the SED-mode data.

\subsection{PDS Luminosity and Mass}

Our primary reason for calculating the stellar and IR excess luminosity
is to measure the fractional luminosities of those
stars that show direct evidence of PDS formation and
to provide well defined upper limits for those stars
that fail to show an excess.
Table~4 lists the stellar luminosity, $L_*\/$,
derived by numerical integration, and the fractional
luminosity, $L_{IR}/L_*\/$, where $L_{IR}\/$ is the luminosity of the
infrared excess that is the result of numerically integrating the flux
density in excess of the predicted level of the photosphere.
Direct measurements of the distance to TWA members are limited to
only five systems (Table~1).
We have assigned a distance of 55~pc, the distance corresponding
to the unweighted average of the parallaxes measured by {\sl Hipparcos\/}
excluding TWA~19, to the other stars in the sample.
In general, $L_*\/$ is higher than expected for the TWA given their
spectral types because
these young, rapidly evolving stars have higher luminosities,
lower temperatures, and larger radii than
when they reach the main sequence.

Conservative upper limits for $L_{IR}/L_*\/$ have been assigned to
objects listed in Table~4 that show no evidence
for a 24~$\mu\/$m excess by assuming that
the 1$\sigma\/$ upper limits estimated for
the 70~$\mu\/$m flux densities represent the brightness
of a possible IR excess plus photosphere at
this wavelength.
It is also assumed
that
any possible IR excess
is well approximated by a blackbody with temperature, $T_{PDS}\/$.
The upper limit on the temperature of a possible cold PDS is
given in Table~4 and
is set so that 
$T_{PDS}\/$ and the assumed value for
$F\/$(70~$\mu\/$m) raise $F\/$(24~$\mu\/$m) by no more than 10\%; the adopted 
uncertainty in the absolute flux calibration of the MIPS
measurements at 24~$\mu\/$m. 
These upper limits on temperature do not preclude small amounts of warmer dust 
within systems that are still consistent with the 24~$\mu\/$m measurements,
but they do constrain the properties of circumstellar material able
to produce substantial 
excesses at 70~$\mu\/$m and be undetected by {\it Spitzer\/}.

The mass of circumstellar dust responsible for the luminosity of
IR excesses observed
in the TWA, or that could produce the upper limits found for $T_{PDS}\/$
and $L_{IR}/L_*\/$, can be deduced following the 
inferences of \citet{jura95} and \citet{chen01}.
That is, the minimum mass in dust can be estimated by
$$M_d \geq (16/3) \pi (L_{\rm IR}/L_*) \rho R^2_{PDS} \langle a \rangle$$
\citep{jura95}, 
where $\langle a \rangle$ is the average radius of a dust grain, 
$R_{PDS}\/$ is the minimum distance from the star where the grains are
in radiative equilibrium, and $\rho\/$ is the mass density of the grains.
The equation above assumes that the dust is radiating from a 
thin shell at a distance $R_{PDS}\/$ from the illuminating star and that
the debris disk is optically thin.
Also, it is assumed that the dust grains are spherical and
have cross sections equal to their geometric cross sections.
In Table~4, we tabulate $M_d\/$ using the values for average grain size and
density adopted by \citet{chen01} ($\langle a \rangle = 2.8$~$\mu$m;
$\rho = 2.5$~g~cm$^{-3}$).
The minimum dust mass responsible for detected IR excesses range from
$\sim 10^{25}$~g for HR~4796A to $\sim 10^{23}$~g for TWA~13.
For HD~98800B, the assumption that the debris disk is optically
thin is not valid, accounting for the much smaller mass estimate
than that determined by \citet{low99}.

The mass estimates in Table~4 for the TWA are conservative lower bounds
for the total circumstellar dust mass in these systems since the
MIPS observations are not sensitive
to colder material located further
from a star and emission from larger particles.
For example, there is evidence for material around
TWA~7 colder than the $\sim$80~K dust detected by {\it Spitzer\/}.
Observations at 850~$\mu\/$m by \citet{webb00a} detected TWA~7
with $F_\nu \sim 15$~mJy.
This flux density is about a factor of seven higher than the extrapolation
to 850~$\mu\/$m of the $\sim$80~K blackbody determined by the 24
and 70~$\mu\/$m excesses.
\citet{webb00a} and \citet{webb00b} interpret
the sub-millimeter emission as coming from 
20~K dust with grain sizes on the order of a few hundred microns.
As a consequence, the mass estimate for the dust disk around TWA~7
derived by the sub-millimeter measurement
is $\sim 10^3 \times\/$ greater than that estimated from the IR excess.

\section{Discussion}

\subsection{Infrared Properties of the TWA}

The main result of the {\it Spitzer\/}-MIPS observations of the TWA is to
show that most stars in the association exhibit no evidence for
circumstellar dust and that the 24~$\mu\/$m data
place severe limits on the presence of warm ($T \gtrsim 100$~K) dust within
these systems.
Although the TWA includes four remarkable IR-excess systems,
there are no other objects that come close to their 24~$\mu\/$m output relative
to the observed near-IR photospheric emission.
The bimodal distribution of warm dust in the
TWA, identified by \citet{weinberger04}
at shorter wavelengths, is confirmed at 24~$\mu\/$m for a
larger sample of objects
(see Figure~1).

The bimodal nature of 70~$\mu\/$m excesses is not as pronounced as at
shorter wavelengths, but there is still
a gap of about a factor of 10 between the strengths of the 70~$\mu\/$m
excess of Hen~3-600 and the {\it Spitzer\/} detections of TWA~7 and 13. 
There is also a range of at least $\sim 300$ between the brightest and 
faintest of the 70~$\mu\/$m excesses in the TWA relative to their 
photospheres at this wavelength.
The luminosity of the observed IR excesses range from $\sim 25\/$\% (TW~Hya and 
HD~98800B) to $\sim 0.1\/$\% (TWA~13) of $L_*\/$.

It is difficult to fully characterize the IR excesses of TWA~7 and 13 given
that they only begin to appear at $\sim$24~$\mu\/$m.
In the case of TWA~13, there is also the complicating factor that
the system is a binary and the distribution of the material responsible for
the excess is unknown.
However, calculating a color temperature for the two objects yields
similar results: $T_{\rm c} \sim 60$--80~K.
Assuming that the excesses are well fit by blackbodies, both systems
have a fractional IR luminosity of
$\sim$1\% of that measured for HD~98800B;
$L_{IR}/L_{*} = 2.0\times 10^{-3}$ and
$1.7\times 10^{-3}$, for TWA~7 and 13, respectively.
Although the lack of observed excess 24~$\mu\/$m emission constrains
the temperatures of possible PDSs for most of the 70~$\mu\/$m
non-detections to be $< 100$~K, the MIPS observational upper limits
cannot rule out IR excesses with fractional luminosity in the 
range of $10^{-3}$--$10^{-4}$.
Even at the estimated young age of the association
it may already be more likely to detect cooler material,
further away from the stars, than warmer material that results
in an excess at 24~$\mu\/$m or shorter wavelengths.
Clearly, at 70~$\mu\/$m wavelength, large amounts of additional
{\it Spitzer\/} time will be needed to detect and
measure the cool dust and debris around these stars.

The variety of the IR properties in the TWA is remarkable
given the assumption that its members have similar ages.
At 8--10~Myr after the formation of the TWA, there are still two systems
(TW~Hya and Hen~3-600)
accreting presumably primordial dust and they
have enough circumstellar material to reprocess a large
fraction (10--30\%) of the
stellar luminosity into the infrared.
Also in the association are two multiple-star
systems, HD~98800 and HR~4796, that 
have a prominent PDS around only one of the stellar
components and the debris is relatively warm ($T_{PDS} = 100$--170~K). 
TWA~7 shows evidence for a weak 24~$\mu\/$m excess with the
detected material
being cooler ($T_{PDS} \sim 80$~K)
and located in orbits
$\gtrsim 7$~AU from the star.
The remaining 19 stars, including TWA~7 and 13, have apparently been cleansed
to a high degree of 
dust with $T > 100$~K.
This variety of properties suggests that dust is swept from regions
within several AU (see Table~4) of the star
very early in the system's evolution.
The fact that stars in the association show no signature of warm dust while
other members exhibit strong T~Tauri activity implies that the 
transition period from the T~Tauri stage to the more quiescent state
observed for most of the TWA K- and M-type stars, is very short.

An efficient mechanism for the removal of dust is thought to be the
formation of planets that sweep up dust and larger debris 
in the regions that they orbit.
\citet{weinberger04} suggest that planetary formation 
was rapidly completed in the TWA at least in
the region where terrestrial planets
would be expected to form.
This led to the rapid disappearance of dust in the inner several AU 
of these systems, leaving only the stellar photosphere to be detected
at wavelengths of 24~$\mu\/$m and shorter.
Strong stellar winds during the T~Tauri stage and the Poynting-Robertson
drag are also mechanisms that can decrease the IR emission
from a system by ridding the environment of small dust grains.
Of course, in most TWA systems, and within 10~Myr of their formation,
dust destruction mechanisms must more than compensate
for the dust production mechanism of collisions between larger bodies to
clear out the terrestrial planet region. 
HD~98800B and HR~4796A are likely cases where their debris systems
indicate that collisions between bodies are still important and produce
dust close enough to the stars to result in large
excesses at 24~$\mu\/$m.

Nearly all members of the observed sample are X-ray emitters, supporting
the assertion that stellar winds in these young, low-mass stars
may be an important dust destruction mechanism
\citep[see e.g.,][]{hollenbach00}.
We list the X-ray--to--stellar luminosity ratios, $L_X\//L_*\/$,
for TWA stars in the last column of Table~4.
The X-ray data are drawn from the {\sl ROSAT\/} All Sky Survey Catalog
and $L_X\/$ is calculated in the manner described by \citet{sterzik99}.
\citet{chen05} find an apparent correlation between IR and X-ray
luminosity for their sample of F- and G-type stars with ages spanning 5--20~Myr
in the
Scorpius-Centaurus OB Association.
This possible correlation is in the sense of brighter 24~${\mu}$m excesses
tending to be associated with fainter stars in X-ray flux, and \citet{chen05}
suggest that the stronger stellar wind implied by the
chromospheric activity
may help explain
the general lack of young systems with 24~${\mu}$m excesses.
Given that only three stars in the TWA sample have
confirmed PDSs that produce measurable excesses at 24~${\mu}$m,
it is not possible to test the validity of the X-ray--IR flux correlation.
However, HD~98800B, HR~4796A, and TWA~7, are all underluminous in X-rays
compared to the average observed X-ray emission for the TWA.

\subsection{Comparison of the TWA with the Young Solar System}

We can compare the {\it Spitzer\/}
observations of young stars with models for the early
Solar System. 
Following the discussions in \citet{gaidos99} and \citet{jura04}, we assume
that the rate of dust production directly scales as the rate of lunar
cratering during the Late Heavy Bombardment.
The rate of dust production, ${\dot M}_{dust}$,
is given by:
$$
{\dot M}_{dust}\;=\;{\dot M}_{0}\,\left(1\,+\,{\beta}\,e^{\frac{t_{9}}{\tau}}\right),
$$
where $t_{9}$ is the look-back time in 10$^9$~yr (Gyr), ${\dot M}_{0}$ is
the current rate of dust production and ${\beta\/}$ and ${\tau\/}$ are fitting
constants such that ${\beta} = 1.6 {\times} 10^{-10}$ and ${\tau}$ = 0.144
\citep{chyba91}.
In this model, the dust production rate has been approximately constant during
the past 3.3~Gyr, but
was as high as
$\sim 10^{4}$ the current value when $t_{9} = 4.6$.
We set ${\dot M}_{0} = 3 {\times} 10^{6}$~g~s$^{-1}$ based on
the zodiacal light in the Solar System \citep{fixsen02}, implying
that the early Solar System
may have had a dust production rate of $3 {\times} 10^{10}$~g~s$^{-1}$.  
In a model where this quantity of dust is produced far from the star and then
loses angular momentum under the action of the Poynting-Robertson effect,
it is straightforward to show \citep[e.g.,][]{gaidos99} that the luminosity of
the dust, $L_{\nu}$, is given by the expression:
$$
L_{\nu}\;=\;\frac{{\dot M}_{dust} c^{2}}{{\nu}}.
$$

For the TWA, young M-type stars would have luminosities at 24~${\mu}$m
of $2 {\times} 10^{18}$~erg~s$^{-1}$~Hz$^{-1}$ and, for distances
of $\sim$50~pc, we expect F$_{\nu}$(24 ${\mu}$m) $\approx$ 700~mJy, vastly
greater than what is observed.
This result suggests the absence of terrestrial planet-forming
environments around these stars.  
However, in M-type stars where the luminosity is relatively low and the stellar
wind rate is relatively large, dust grains mainly lose angular momentum by 
stellar wind drag rather than through Poynting-Robertson
drag \citep[see][]{jura04,plavchan05}.
In this case, and if ${\dot M}_{wind}$ denotes the stellar wind loss rate, then:
$$
L_{\nu}\;=\;\frac{{\dot M}_{dust}}{{\dot M}_{wind}}\,\frac{L_{*}}{{\nu}}.
$$
Although we do not know ${\dot M}_{wind}$ for young M-type stars, we can
extrapolate from their X-ray emission since the winds are likely to be driven
by the same hot corona which produces the X-rays.
It is  plausible that
young M-type stars have mass loss rates 10$^{2}$ greater than the current
Solar wind loss rate of $2 {\times} 10^{12}$~g~s$^{-1}$, and
we estimate that ${\dot M}_{wind} = 2 {\times} 10^{14}$~g~s$^{-1}$.
For main sequence early M-type stars, we may adopt $L_{*} = 0.05 L_{\odot}$.
Therefore, we expect that at 24~${\mu}$m, 
$L_{\nu} = 2 {\times} 10^{15}$~erg~s$^{-1}$~Hz$^{-1}$ and
a predicted flux at 50~pc of only $\sim$0.7~mJy.
A 24~${\mu}$m excess this small is beyond the sensitivity of our
current measurements,
and it is possible that planet forming activity may still
be occurring around the young
M-type stars in the TWA even though we do not detect an infrared
excess.

The TWA gives us an important
example of the rapid evolution of the circumstellar
material around generally low-mass stars with ages of $\sim 10$~Myr that
is in line with time scales of terrestrial planet formation inferred
for our solar system from radiochemistry of meteorites \citep{yin02,kleine02}.
The diverse IR properties of the association also suggest that the 
evolutionary processes relevant to planetary formation and accretion disk
dispersal
may occur at different rates
even for stars with similar spectral types.
For most of the systems observed, the lack of emission from dust within a
few AU of the stars leads to two possible conclusions.
Either, (1) the early conditions around pre-main sequence M and late K stars
prevent the formation of terrestrial planets in most cases,
or that (2) terrestrial planet building
has already progressed to the point where
the mass is in the form of planetesimals.

\acknowledgments

This work is based on observations made with the
{\it Spitzer Space Telescope\/}, which is
operated by the Jet Propulsion Laboratory (JPL), California Institute of
Technology (CIT), under National Aeronautics and Space Administration (NASA) 
contract 1407.
We thank NASA,
JPL, and the {\it Spitzer\/} Science Center for 
support through {\it Spitzer\/}, MIPS,
and Science Working Group contracts 960785,
959969, and 1256424 to The University of Arizona.
We thank M. Blaylock, C. Engelbracht,
K. Gordon, K. Misselt, J. Muzerolle, G. Neugebauer,
J. Stansberry, K. Stapelfeldt,
K. Su, B. Zuckerman, and an anonymous referee
for useful comments and discussions.
V. Krause acknowledges a Summer Undergraduate Research Fellowship at JPL.
This publication makes use of data products from the Two-Micron All Sky
Survey, which is a joint project of the University of Massachusetts
and the Infrared Processing and Analysis
Center/CIT,
funded by NASA
and the National Science Foundation.

\clearpage
\vskip 1.0in
\begin{deluxetable}{lllccrccc}
\tablecolumns{9}
\tablewidth{0pc}
\tabletypesize{\footnotesize}
\tablecaption{The TWA Sample}
\tablehead{
\colhead{Star\tablenotemark{a}} &
\colhead{Other Name} & \colhead{Spectral Type} &
\colhead{Ref.\tablenotemark{b}} &
\colhead{distance\tablenotemark{c}} &
\colhead{$V\/$} &
\colhead{Ref.\tablenotemark{d}} & \colhead{$K_s\/$\tablenotemark{e}} &
\colhead{$T_*\/$\tablenotemark{f}} \\
\colhead{} & \colhead{} & \colhead{} & \colhead{} &
\colhead{(pc)} &
\colhead{} & \colhead{} & \colhead{} &
\colhead{(K)} }

\startdata

TWA~1 & TW~Hya & K8e & 1 & 56\ $\pm$\ 7 & 11.07 & 1--9 & 7.30 & 3700 \\
TWA~2AB & CD$-$29$^{\circ}$8887 & M2e+M2 & 2 & \nodata & 11.42 & 1--4,6,8,9 & 6.71 & 3500 \\
TWA~3AB & Hen~3-600 & M3e+M3.5 & 2 & \nodata & 12.04 & 1,2,6,7 & 6.77 & 3300 \\
TWA~4AB & HD~98800 & K4+K5 & 3 & 47\ $\pm$\ 6 & 8.89 & 1--7, 10--15 & 5.59(A), 5.90(B) & 4800(A), 3800(B) \\
TWA~5A & CD$-$33$^{\circ}$7795 & M3e & 3 & \nodata & 11.72 & 1--5,8,9,16 & 6.75 & 3400 \\
 & & & & & & & & \\
TWA~6 & & K7 & 4 & \nodata & 12.00 & 1--4,6,12,16,17 & 8.04 & 4000 \\
TWA~7 & & M1 & 4 & \nodata & 11.06 & 1--4,6,9,16 & 6.90 & 3500 \\
TWA~8A & & M2e & 4 & \nodata & 13.30 & 1--3,6,12 & 7.43 & 3300 \\
TWA~8B & & M5 & 4 & \nodata & & & 9.01 & 3100 \\
TWA~9A & CD$-$36$^{\circ}$7429A & K5 & 4 & 50\ $\pm$\ 6\tablenotemark{g} & 11.32 & 1--6,9,12 & 7.85 & 4200 \\
TWA~9B & CD$-$36$^{\circ}$7429B & M1 & 4 & 50\ $\pm$\ 6\tablenotemark{g} & 14.10 & & 9.15 & 3300 \\
TWA~10 & & M2.5 & 4 & \nodata & 12.70 & 1,2,4,6,12 & 8.19 & 3400 \\
 & & & & & & & & \\
TWA~11A & HR~4796A & A0 & 5 & 67\ $\pm$\ 2 & 5.78 & 2--4,6--9,12 & 5.77 & 9250 \\
TWA~12 & & M2 & 6 & \nodata & 13.60 & 2,4,12,16 & 8.06 & 3000 \\
TWA~13A & CD$-$34$^\circ$7390A & M1e & 1,6 & \nodata & 12.10 & 1,4,12,16 & 7.50 & 3300 \\
TWA~13B & CD$-$34$^\circ$7390B & M1e & 1,6 & \nodata & 12.40 & & 7.46 & 3200 \\
TWA~14 & & M0 & 7 & \nodata & 13.80 & 2--4,12,16,18 & 8.50 & 3300 \\
TWA~15A & & M1.5 & 7 & \nodata & 14.10 & 2,4,16,18 & 9.68 & 3500 \\
TWA~15B & & M2 & 7 & \nodata & 14.00 & & 9.56 & 3500 \\
 & & & & & & & & \\
TWA~16 & & M1.5 & 7 & \nodata & 12.30 & 2,4,12,16,18 & 8.10 & 3700 \\
TWA~17 & & K5 & 7 & \nodata & 12.70 & 2,4,5,12,16,18 & 8.98 & 4000 \\
TWA~18 & & M0.5 & 7 & \nodata & 12.90 & 2,4,5,12,16,18 & 8.84 & 3800 \\
TWA~19A & HD~102458A & G5 & 7 & 104\ $\pm$\ 15 & 9.10 & 2--5,7,12,16 & 7.51 & 6000 \\
TWA~19B & HD~102458B & K7 & 7 & 104\ $\pm$\ 15 & 11.90 & 2,4,12,16,18 & 8.28 & 4000 \\
\enddata

\tablenotetext{a}{AB designates binaries that are unresolved by
{\it Spitzer\/} at $\lambda = 24~\mu\/$m.}
\tablenotetext{b}{References for spectral types and
identifications as members of the TWA ---
(1) \citet{rucinski83}; (2) \citet{reza89}; (3) \citet{hetem92};
(4) \citet{webb99}; (5) \citet{houk82}; (6) \citet{sterzik99}; 
(7) \citet{zuckerman01}.}
\tablenotetext{c}{Data are from the {\sl Hipparcos\/} Catalog
\citep{perryman97}.}
\tablenotetext{d}{Sources and references for previous optical--IR photometry ---
(1) \citet{torres00}; (2) \citet{reid03};
(3) {\sl Hipparcos\/}-Tycho \citep{ESA97};
(4) 2MASS \citep{2MASS}; (5) DENIS \citep{DENIS}; (6) \citet{jayawardhana99};
(7) {\sl IRAS\/} \citep{IRAS};
(8) UCAC1 \citep{zacharias00}; (9) Yale-SPM \citep{platais98};
(10) \citet{low99}; (11) UCAC2 \citep{zacharias04}; (12) GSC \citep{lasker90};
(13) \citet{soderblom98}; (14) \citet{prato01}; (15) \citet{gehrz99};
(16) \citet{weinberger04}; (17) USNO A2.0 \citep{monet99};
(18) \citet{zuckerman01}.}
\tablenotetext{e}{Apparent $K_s\/$ magnitude is from the 2MASS
Point Source Catalog, except for HD~98800, where
the $K\/$-band magnitudes for both components are from \citet{low99}.}
\tablenotetext{f}{Temperature of the best-fit Kurucz or NextGen model to the
stellar photometry.}
\tablenotetext{g}{See \citet{mamajek00} for a discussion of the
{\sl Hipparcos\/} parallax measurement for TWA~9.}

\end{deluxetable}

\clearpage

\begin{deluxetable}{lrrrccc}
\tablecolumns{7}
\tablewidth{0pc}
\tabletypesize{\small}
\tablecaption{MIPS Bandpasses}
\tablehead{
\colhead{Band}  &
\colhead{$\lambda_0\/$} &
\colhead{$\Delta\lambda\/$} &
\colhead{scale} &
\colhead{$C_\nu\/$\tablenotemark{a}} &
\colhead{Phot. Apt.} &
\colhead{$A_\nu\/$\tablenotemark{b}} \\
\colhead{} & 
\colhead{($\mu\/$m)} & 
\colhead{($\mu\/$m)} & 
\colhead{(\arcsec~pixel$^{-1}$)} & 
\colhead{} & 
\colhead{(radius in~\arcsec )} & 
\colhead{}}
\startdata

24~$\mu\/$m & 23.7 & 4.7 & 2.55 & 1.042 $\times$ 10$^{-3}$ & 15 or 37.5 & 1.15 or 1.06 \\
70~$\mu\/$m wide field & 71 & 19 & 9.85 & 14.9  & 29.6 & 1.35 \\
70~$\mu\/$m narrow field & 71 & 19 & 5.24 & 52.7 & 29.7 & 1.35 \\
160~$\mu\/$m & 156 & 35 & 16 & 1.0 & 64 & 1.33 \\

\enddata

\tablenotetext{a}{Conversion factor from the calibrated instrumental units
of the DAT output to mJy.}
\tablenotetext{b}{Factor to correct aperture photometry to that expected from
a hypothetical aperture of infinite extent.}

\end{deluxetable}

\begin{deluxetable}{lrrcrcrc}
\tablecolumns{8}
\tablewidth{0pc}
\setlength{\tabcolsep}{0.02in}
\tabletypesize{\small}
\tablecaption{MIPS Observations}
\tablehead{
\colhead{Star}  &
\colhead{Epoch} &
\colhead{24~$\mu\/$m} &
\colhead{(S/N)$_{24~\mu\/m}$} &
\colhead{70~$\mu\/$m\tablenotemark{a}} &
\colhead{(S/N)$_{70~\mu\/m}$} &
\colhead{160~$\mu\/$m} &
\colhead{(S/N)$_{160~\mu\/m}$} \\
\colhead{} & 
\colhead{(JD-2453000.0)} & 
\colhead{(mJy)} & 
\colhead{} & 
\colhead{(mJy)} & 
\colhead{} & 
\colhead{(mJy)} & 
\colhead{}}
\startdata

TW~Hya & 38.282 & 2270\ $\pm$\ 230 & $>$ 10$^4$ & 3640\ $\pm$\ 730 & 170 & 6570\ $\pm$\ 1310 & 160 \\
TWA~2AB & 35.384 & 20.1\ $\pm$\ 2.0 & 200 & $<$ 40 & \nodata & \nodata & \nodata \\
Hen 3-600 & 38.268 & 1650\ $\pm$\ 170 & $>$ 10$^4$ & 700\ $\pm$\ 140 & 140 & 740 \ $\pm$\ 190 & 6 \\
HD 98800AB & 177.665 & 8500\ $\pm$\ 350 & 1000 & 6260\ $\pm$\ 1250 & 200 & 2100\ $\pm$ 420 & 67 \\
TWA~5A & 38.255 & 21.1\ $\pm$\ 2.1 & 220 & $<$ 9 & \nodata & \nodata & \nodata\\
 & & & & & & & \\
TWA~6 & 134.825 & 5.7\ $\pm$\ 0.6 & 240 & $<$ 8 & \nodata & \nodata & \nodata \\
TWA~7 & 35.398 & 30.2\ $\pm$\ 3.0 & 260 & 85\ $\pm$\ 17 & 22 & \nodata & \nodata \\
TWA~8 & 35.905 &  14.5\ $\pm$\ 1.5 & 160 & $<$ 11 & \nodata & \nodata & \nodata \\
\hfil A & & 11.1\ $\pm$\ 1.5 & & & & & \\
\hfil B & & 3.4\ $\pm$\ 0.4 & & & & & \\
TWA~9 & 37.870 & 9.3\ $\pm$\ 1.1 & 160 & $<$ 34 & \nodata & \nodata & \nodata \\
\hfil A & & 6.0\ $\pm$\ 0.6 & & & & & \\
\hfil B & & 3.3\ $\pm$\ 0.3 & & & & & \\
TWA~10 & 55.726 & 5.1\ $\pm$\ 0.5 & 190 & $<$ 8 & \nodata & \nodata & \nodata
\\
 & & & & & & & \\
HR~4796A & 35.928 & 3030\ $\pm$\ 303 & $>$ 10$^4$ & 5160\ $\pm$\ 1100 & 240 & 1800\ $\pm$\ 360 & 55 \\
TWA~12 & 177.907 & 5.7\ $\pm$\ 0.6 & 50 & $<$ 29 & \nodata & \nodata & \nodata \\
TWA~13 & 177.680 & 17.6\ $\pm$\ 1.8 & 150 & 27.6\ $\pm$\ 5.9 & 13 & \nodata & \nodata \\
\hfil A & & 8.8\ $\pm$\ 1.3  & & & & & \\
\hfil B & & 8.8\ $\pm$\ 1.3  & & & & & \\
TWA~14 & 55.677 & 4.1\ $\pm$\ 0.4 & 27 & $<$ 25 & \nodata & \nodata & \nodata \\
TWA~15 & 55.701 & 2.9\ $\pm$\ 0.3 & 72 & $<$ 12 & \nodata & \nodata & \nodata \\
\hfil A & & 1.4\ $\pm$\ 0.2 & & & & & \\
\hfil B & & 1.5\ $\pm$\ 0.2 & & & & & \\
 & & & & & & & \\
TWA~16 & 55.714 & 6.6\ $\pm$\ 0.6 & 45 & $<$ 18 & \nodata & \nodata & \nodata \\
TWA~17 & 55.743 & 1.5\ $\pm$\ 0.2 & 13 & $<$ 13 & \nodata & \nodata & \nodata \\
TWA~18 & 55.762 & 2.3\ $\pm$\ 0.3 & 20 & $<$ 9 & \nodata & \nodata & \nodata \\
TWA~19A & 55.689 & 10.4\ $\pm$\ 1.1 & 100 & $<$ 27 & \nodata & \nodata & \nodata \\
TWA~19B & 55.689 & 4.6\ $\pm$\ 0.5 & 44 & $<$ 27 & \nodata & \nodata & \nodata \\

\enddata

\tablenotetext{a}{Upper limits listed are the detected signal
within the photometric aperture plus 3$\sigma\/$. If the flux measured within
the aperture is less than the background level measured for the
aperture, the upper limit listed is 3$\sigma\/$.}

\end{deluxetable}

\clearpage

\begin{deluxetable}{lrcccrc}
\tablecolumns{7}
\tablewidth{0pc}
\tabletypesize{\small}
\tablecaption{Properties of Dust Disks in the TWA}
\tablehead{
\colhead{Star}  &
\colhead{$T_{PDS}\/$ (K)} &
\colhead{$R_{PDS}\/$\tablenotemark{a} (AU)} &
\colhead{$M_{d}\/$ ($10^{23}$ g)\tablenotemark{b}} &
\colhead{$L_*\//L_{\sun}\/$\tablenotemark{c}} &
\colhead{$L_{IR}/L_*\/$} &
\colhead{$L_{X}/L_*\/$}}
\startdata

TW Hya & \nodata & \nodata & \nodata & 0.24 & $2.7 \times 10^{-1}$ & $2.7 \times 10^{-3}$ \\
TWA 2AB\tablenotemark{d} & $< 85$ & $> 7.0$ & 0.27 & 0.39 & $< 2.1 \times 10^{-4}$ & $5.9 \times 10^{-4}$ \\
Hen 3-600\tablenotemark{d} & \nodata & \nodata & \nodata & 0.35 & $1.2 \times 10^{-1}$ & $6.1 \times 10^{-4}$ \\
HD~98800B\tablenotemark{d} & 160 & 2.2 & $> 29$\tablenotemark{e} & 0.53 & $2.2 \times 10^{-1}$ & $7.1 \times 10^{-4}$ \\
 & & & & \\
TWA 5A & $< 145$ & $> 2.3$ & 0.01 & 0.36 & $< 9.6 \times 10^{-5}$ & $1.1 \times 10^{-3}$ \\
TWA 6 & $< 110$ & $> 2.6 $ & 0.02 & 0.14 & $< 1.4 \times 10^{-4}$ & $1.4 \times 10^{-3}$ \\
TWA 7 & $80$ & 6.8 & 2.4 & 0.31 & $2.0 \times 10^{-3}$ & $7.8 \times 10^{-4}$ \\
TWA 8A & $< 150$ & $> 1.5$ & 0.04 & 0.19 & $< 6.6 \times 10^{-4}$ & $1.3 \times 10^{-3}$ \\
TWA 8B & $< 70$ & $> 3.2$ & 0.19 & 0.04 & $< 7.2 \times 10^{-4}$ & \nodata\hfil \\
 & & & & \\
TWA 9A & $< 60$ & $> 8.6$ & 1.3 & 0.15 & $< 6.8 \times 10^{-4}$ & $1.3 \times 10^{-3}$ \\
TWA 9B & $< 60$ & $> 4.2$ & 1.5 & 0.03 & $< 3.2 \times 10^{-3}$ & \nodata\hfil \\
TWA 10 & $< 80$ & $> 3.7$ & 0.09 & 0.09 & $< 2.6 \times 10^{-4}$ & $1.1 \times 10^{-3}$ \\
HR~4796A & $108$ & 30 & 110 & 19.7 & $4.8 \times 10^{-3}$ & \nodata\hfil \\
TWA 12 & $< 65$ & $> 6.0$ & 0.94 & 0.09 & $< 1.0 \times 10^{-3}$ & $1.0 \times 10^{-3}$ \\
 & & & & \\
TWA 13A & $65$ & 7.8 & 1.4 & 0.18 & $8.6 \times 10^{-4}$ & $1.8 \times 10^{-3}$ \\
TWA 13B & $65$ & 7.7 & 1.4 & 0.17 & $8.9 \times 10^{-4}$ & \nodata \\
TWA 14 & $< 65$ & $> 5.1$ & 0.59 & 0.07 & $< 8.8 \times 10^{-4}$ & $1.4 \times 10^{-3}$ \\
TWA 15A & $< 60$ & $> 3.5$ & 0.27 & 0.03 & $< 8.2 \times 10^{-4}$ & $4.7 \times 10^{-3}$ \\
TWA 15B & $< 65$ & $> 3.1$ & 0.20 & 0.03 & $< 7.9 \times 10^{-4}$ & \nodata\hfil \\
 & & & & \\
TWA 16 & $< 80$ & $> 4.5$ & 0.15 & 0.12 & $< 2.8 \times 10^{-4}$ & $7.0 \times 10^{-4}$ \\
TWA 17 & $< 60$ & $> 5.1$ & 0.54 & 0.06 & $< 7.8 \times 10^{-4}$ & $9.8 \times 10^{-4}$ \\
TWA 18 & $< 80$ & $> 3.1$ & 0.06 & 0.06 & $< 2.6 \times 10^{-4}$ & $9.8 \times 10^{-4}$ \\
TWA 19A & $< 95$ & $> 13$ & 1.0 & 2.10 & $< 2.2 \times 10^{-4}$ & $7.1 \times 10^{-4}$ \\
TWA 19B & $< 70$ & $> 11$ & 2.7 & 0.38 & $< 8.2 \times 10^{-4}$ & \nodata\hfil \\

\enddata

\tablenotetext{a}{The minimum distance from a star for dust in thermal
equilibrium at temperature, $T_{PDS}\/$, is given by
$R_{PDS} = 0.5 (T_*/T_{PDS})^2 R_*\/$, where $R_*\/$
is radius of the star.}
\tablenotetext{b}{The estimated minimum mass
in dust, $M_{d}\/$, consistent 
with the observations for systems where $T_{PDS}\/$ can be 
determined or constrained, and assuming that the circumstellar 
debris is optically thin.
An average dust grain size of
2.8~$\mu\/$m in radius and a density for the material of
$\rho = 2.5$~g~cm$^{-3}$ are assumed \citep{chen01}. 
For systems with no detected IR excesses, $M_{d}\/$ is an estimate using the
limits determined for $R_{PDS}\/$ and $L_{IR}/L_*\/$.}
\tablenotetext{c}{A distance to the TWA of
55~pc is assumed except for the five systems with measured parallaxes. 
For these systems, the distances listed in Table~1 are used to determine 
$L_*\/$.}
\tablenotetext{d}{TWA~2, Hen~3-600, and HD~98800B are binaries and both
stellar components are included in the luminosity values.}
\tablenotetext{e}{The minimum dust mass for the HD~98800B debris 
system is likely to be underestimated because the assumption that the 
circumstellar material is optically thin in the IR is
not valid \citep[see,][]{low99}.}

\end{deluxetable}

\clearpage

\begin{figure}
\figurenum{1}
\vspace{6.0in}
\includegraphics{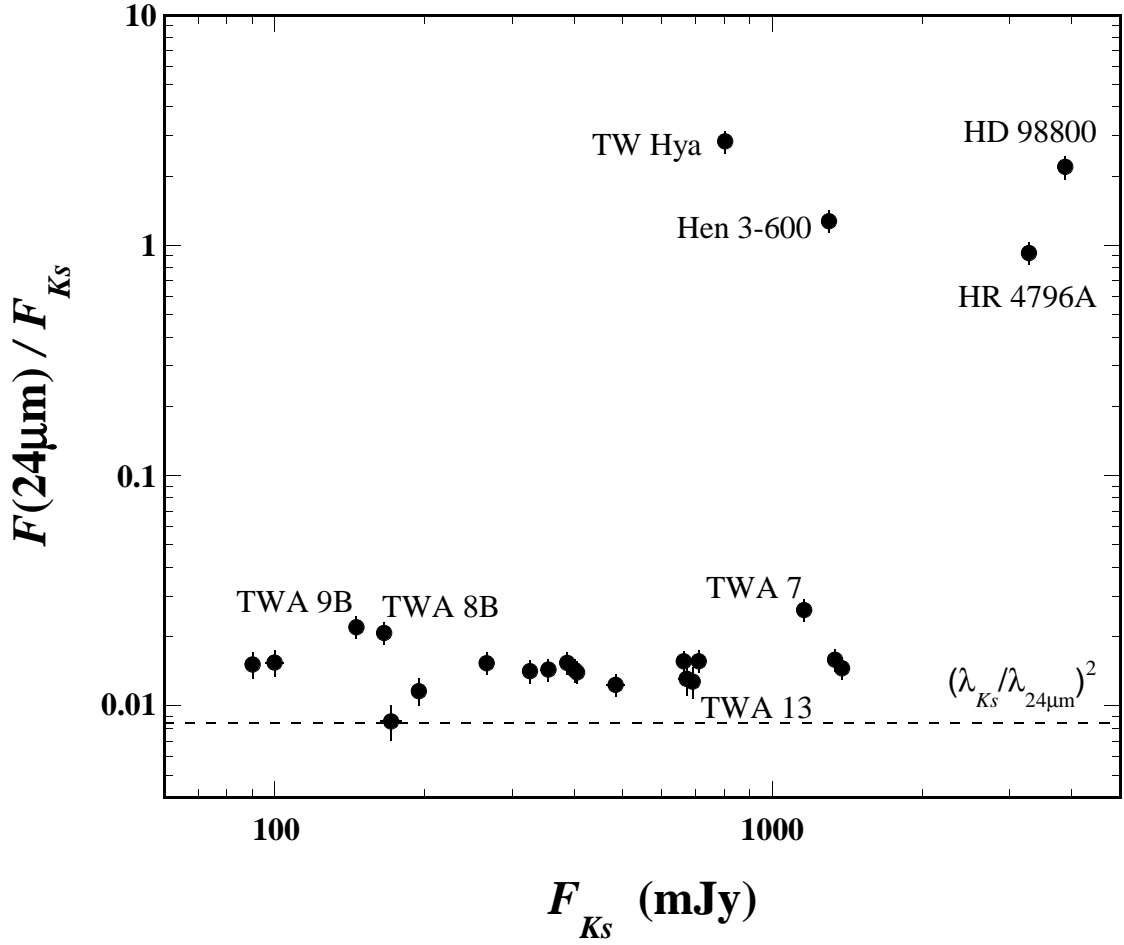}
\caption{MIPS 24~$\mu\/$m photometry compared to the 2MASS $K_s\/$-band
measurements of the TWA sample.
The dashed line represents the $F_{24\mu\/{\rm m}\/}$-to-$F_{Ks}\/$
ratio under the assumption that both photometric bands lie on
the Rayleigh-Jeans tail of the stellar spectrum.}
\label{fig1}
\end{figure}
\clearpage
\begin{figure}
\figurenum{2}
\vspace{6.0in}
\includegraphics{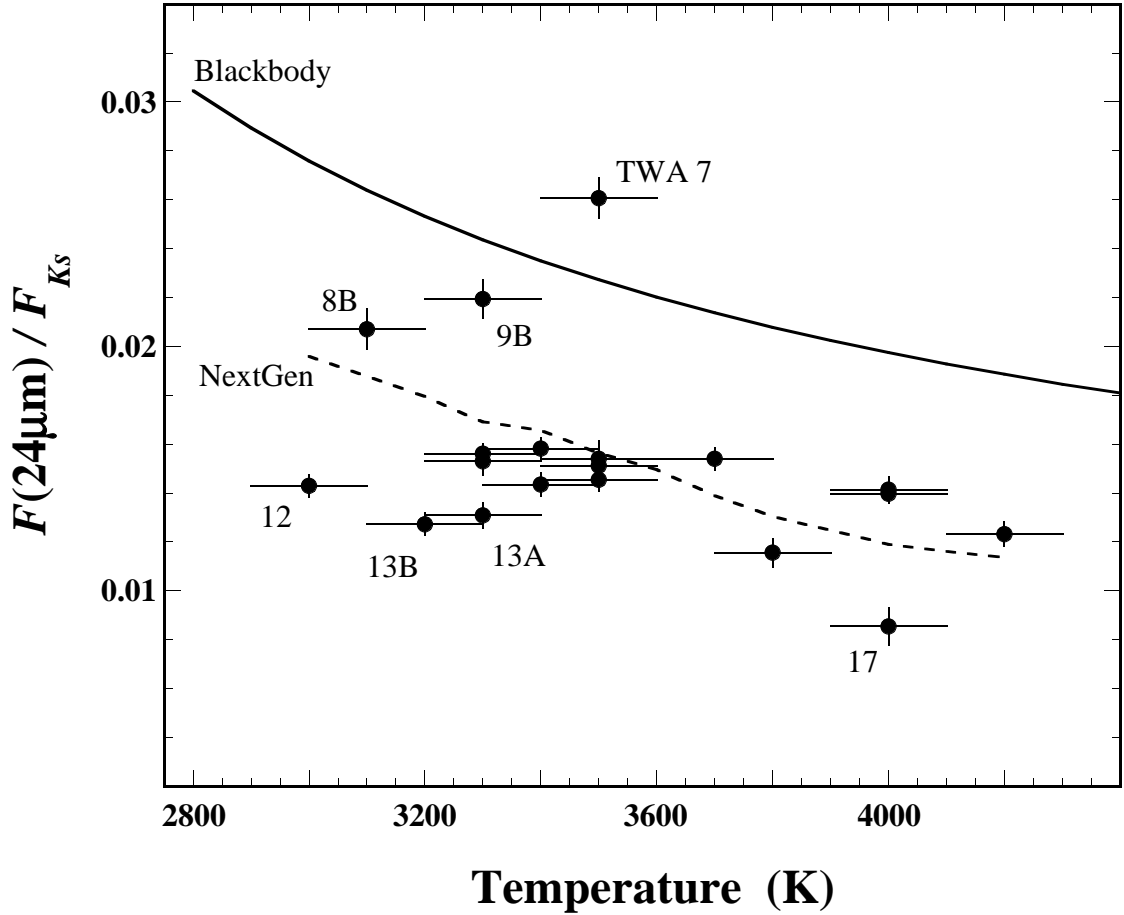}
\caption{The ratio of the observed 24~$\mu\/$m flux density to that determined
for the $K_s\/$-band from the 2MASS Point Source Catalog plotted against
the best-fitting model temperature of the stars.
TW~Hya, Hen~3-600, HD~98800, HR~4796A, and TWA~19A
are not shown so that TWA members
with $T < 6000 K\/$ or without a large 24~$\mu\/$m excess
can be displayed in more detail.
The {\it solid\/} curve represents
the flux ratio between the two filter bandpasses for a blackbody.
Likewise, the {\it dashed\/} curve is the ratio determined
from the NextGen models used to estimate the 24~$\mu\/$m and 70~$\mu\/$m
photospheric flux densities.}
\label{fig2}
\end{figure}
\clearpage

\begin{figure}
\figurenum{3}
\vspace{6.0in}
\includegraphics{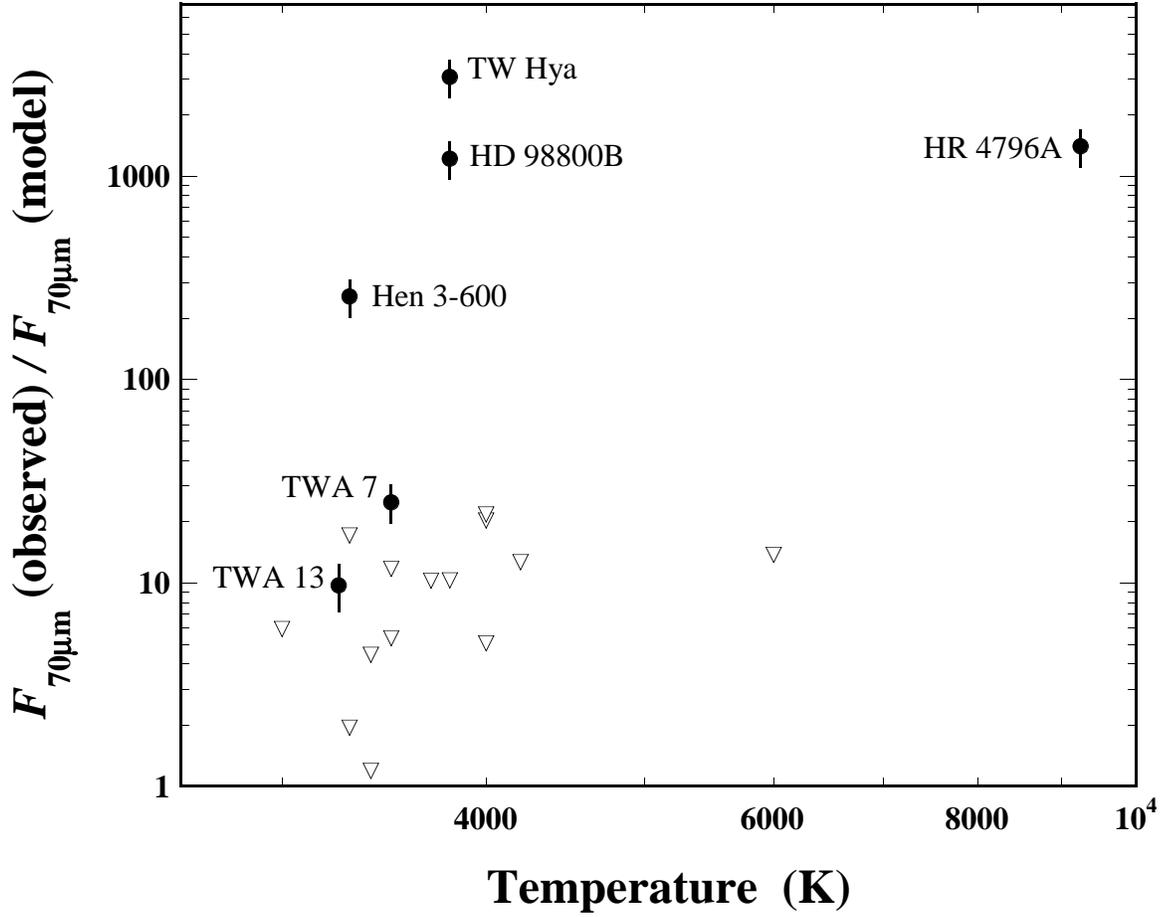}
\caption{The 70~$\mu\/$m excesses measured by {\it Spitzer\/} relative to the
predicted 70~$\mu\/$m flux densities of the stars in the TWA.
As in Figure~2, the data are plotted against the model temperature
of the stellar photospheres.
Each star detected by MIPS at 70~$\mu\/$m is labeled.
One-sigma upper limits are denoted by {\it open triangles\/}.}
\label{fig3}
\end{figure}
\clearpage

\begin{figure}
\vspace{8.2in}
\figurenum{4}
\includegraphics{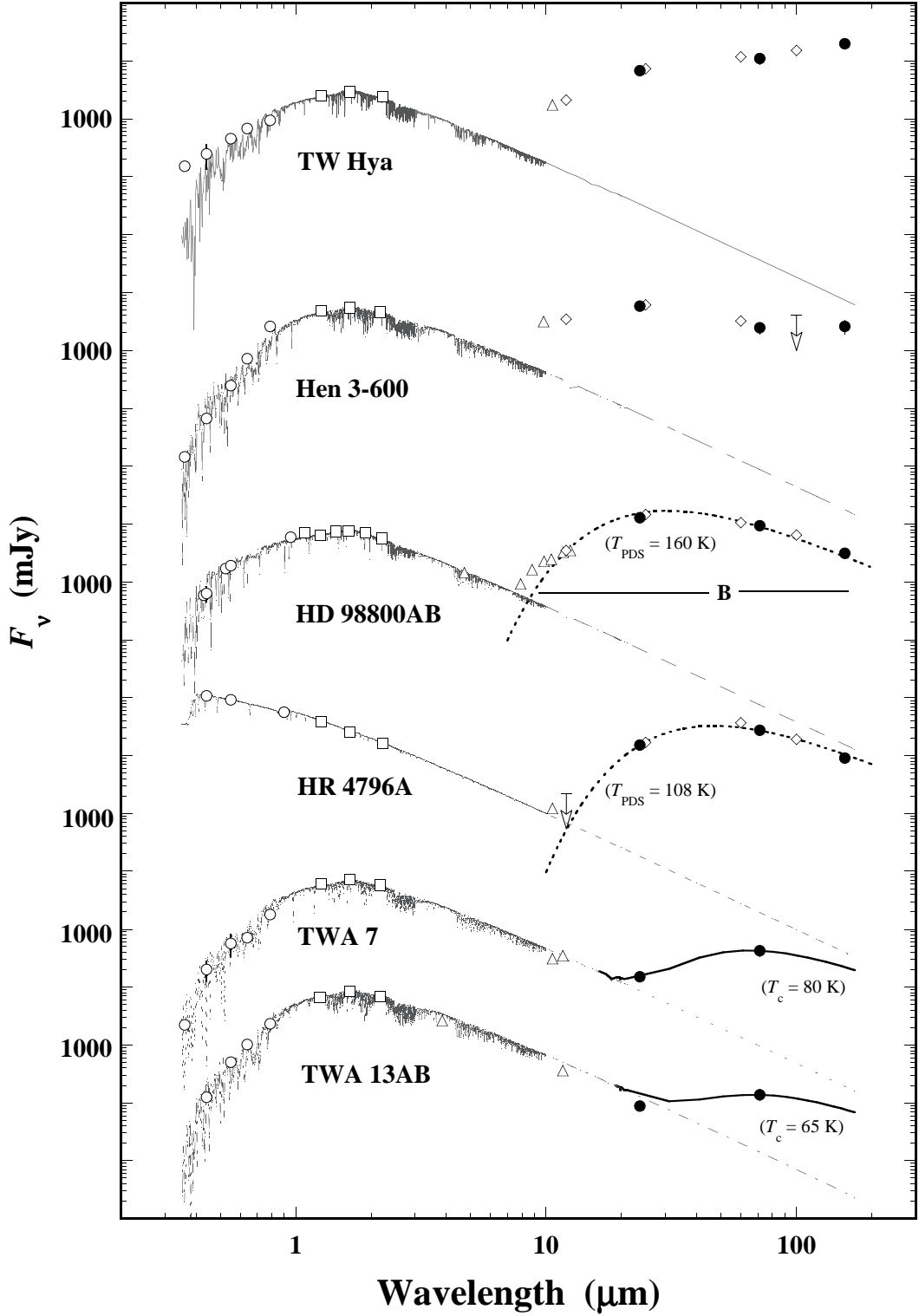}
\caption{The SEDs of the six TWA members with confirmed IR excesses.
Model stellar photospheres are shown for the stars, along with
available photometric measurements from 0.36--160~$\mu\/$m.
Measurements made by MIPS are denoted by {\it filled circles\/}.
Data from the literature are shown by {\it open\/} symbols.
References to the optical ({\it circles\/}), near-IR ({\it squares\/}),
ground-based mid-IR ({\it triangles\/}), and {\sl IRAS\/}
({\it diamonds\/} and upper limits) photometry are given in Table~1.
\citet{gehrz99} showed that HD~98800B produces the
IR excess in this binary system,
though the sum of the models and observed flux densities of
A and B are plotted.
Blackbody fits to the MIPS and {\sl IRAS\/} data for HD~98800B and
HR~4796A are shown as bold {\it dashed\/} curves.
The bold {\it solid\/} curves shown for TWA~7 and 13AB
are the sum of the photospheres and IR excesses
assuming that the excesses are described by blackbodies having
$T_{PDS} = 60$--80~K. 
Color temperatures ($T_{\rm c}$) for TWA~7 and 13AB are also displayed.
The SEDs have been spaced apart for clarity and a 1000~mJy fiducial is
labeled for each star.
To recover the measured flux densities in mJy,
multiply the SEDs by the following factors:
0.33 (TW~Hya), 0.29 (Hen~3-600), 0.67 (HD~98800AB),
and 0.2 (HR~4796A, TWA~7, and 13AB).}
\label{fig4}
\end{figure}
\clearpage

\begin{figure}
\figurenum{5}
\vspace{6.3in}
\includegraphics{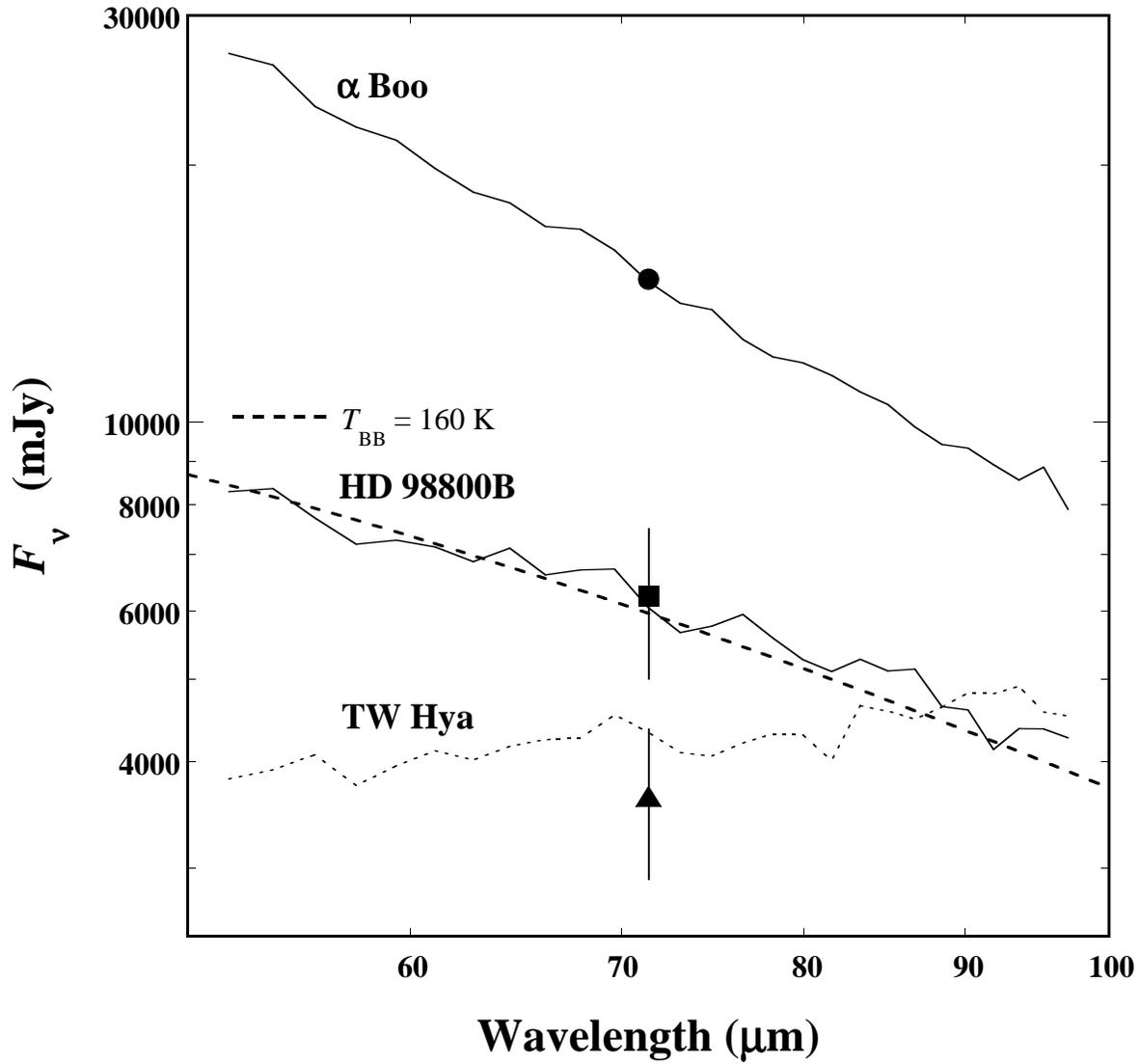}
\caption{MIPS SED-mode spectra of TW~Hya and HD~98800B.  For comparison,
the spectrum of $\alpha\/$~Boo is displayed and was used to calibrate the
fluxes and spectral shapes of the two TWA members.
The broad-band flux densities measured from the MIPS 70~$\mu\/$m photometry
are shown for TW~Hya ({\it square\/}) and HD~98800B ({\it triangle\/}).
The spectra are calibrated assuming that $F_\nu = 14.7$~Jy
at the effective 
wavelength of the 70~$\mu\/$m bandpass for $\alpha\/$~Boo ({\it circle\/}), 
and that its spectrum is adequately described at these wavelengths
by $F_\nu \propto \nu^2\/$.
The best-fit blackbody to the IR excess of HD~98800B using the broad-band
infrared measurements from
{\sl IRAS\/}, {\it Spitzer\/}-MIPS, and the ground is shown by the 
{\it dashed\/} curve and has a temperature, $T_{\rm BB} = 160$~K (see Figure~4).
Most of the apparent ``noise'' in the SED-mode spectra is caused by
pixel-to-pixel
sensitivity variations of the 70~$\mu\/$m array.}
\label{fig5}
\end{figure}

\end{document}